\documentstyle[11pt]{article}
\newcommand{\bm}[1]{\mbox{\boldmath $#1$}}
\begin{document}

\def\baselineskipTaj{0.45cm}
\baselineskip=\baselineskipTaj

\title{
Extensive Hartree-Fock+BCS calculation\\
with Skyrme SIII force
\thanks{UT-Komaba 96-9, to be published in Nuclear Physics A}
}
\author{
Naoki Tajima, Satoshi Takahara, and Naoki Onishi \\
Institute of Physics, College of Arts and Sciences,\\
University of Tokyo, Komaba, Meguro, Tokyo, 153, Japan.
}
\date{}

\maketitle

\begin{abstract}
We have performed deformed Hartree-Fock+BCS calculations with the
Skyrme SIII force for the ground states of even-even nuclei with $2
\leq Z \leq 114$ and $N$ ranging from outside the proton drip line to
beyond the experimental frontier in the neutron-rich side.  We
obtained spatially localized solutions for 1029 nuclei, together with
the second minima for 758 nuclei.  The single-particle wavefunctions
are expressed in a three-dimensional Cartesian-mesh representation,
which is suitable to describe nucleon skins, halos, and exotic shapes
as well as properties of ordinary stable nuclei.  After explaining
some of the practical procedures of the calculations, we compare the
resulting nuclear masses with experimental data and the predictions of
other models. We also discuss the quadrupole ($m$=0, 2) and
hexadecapole ($m$=0, 2, 4) deformations, the skin thicknesses, the
halo radii, and the energy difference between the oblate and the
prolate solutions. Our results can be obtained via computer network.
\end{abstract}

\section{Introduction} \label{s_intro}


It is one of the most important goals of the nuclear theory to
reproduce and predict the nuclear ground-state energy and other
properties globally in the nuclear chart within a single framework.
A variety of theoretical models have been introduced for this goal.
The most elaborate works have been done in the framework of the
finite-range droplet model with a microscopic shell correction (FRDM),
whose latest result was given by M{\"o}ller et al.\ \cite{MNM94}.
Another extensive calculation was carried out by Aboussir et
al.\ \cite{APD92} in the extended Thomas-Fermi plus Strutinsky
integral method (ETFSI).  The former
as well as the latter methods can be regarded as approximations to the
Hartree-Fock (HF) equation.  The straight-forward solutions of the
equation including deformation require long computation time for
global calculations even with present computers.  Such global results
are not yet available to the public as far as we know.  In this paper,
we report on the results of our
extensive HF+BCS calculation with the Skyrme SIII force for 1029
even-even nuclei having atomic number ranging from $2$ to $114$ and
neutron number from outside the proton drip line to beyond the
experimental neutron-rich frontier.


We use a HF+BCS code {\em EV8} \cite{BFH85}, in which the
single-particle wavefunctions are expressed in a three-dimensional
Cartesian-mesh representation, while most of the other methods for
deformed nuclei express the single-particle wavefunctions by the
expansion in a harmonic oscillator basis.  Let us mention three
advantages of the mesh representation.  First, it is capable of
treating nucleon skins and halos. In contrast, they cannot be
described efficiently in the oscillator-basis expansion because the
asymptotic form of wavefunctions far from the nuclear surface is
determined by the basis. Second, one can treat exotic (e.g.,
high-multipole) shapes and large (e.g., super and hyper) deformations
without preparing a basis specific to each shape.  Third, the
saturation property of density makes the atomic nucleus a very
suitable object to the mesh representation, as we discuss in section
\ref{s_basis}.


This paper is mainly concerned with the
proton-rich nuclei than the neutron-rich ones. We thoroughly explore
the proton-rich side, even to beyond the proton drip line by several
neutrons, while restricting the investigations in the neutron-rich
side within the experimental frontier plus a few neutrons.  This is
because the HF+BCS method used in this paper cannot correctly include
the coupling in the pairing channel with the continuum, which can be
influential in nuclei near the neutron drip line\cite{DFT84,DHN94}.
The HF+BCS method including the coupling with the continuum gives rise
to unphysical neutron gas surrounding the nucleus.  The solution of
this problem requires the Hartree-Fock-Bogolyubov method, with which
ordinary and pair densities are spatially localized when the Fermi
level is at a negative energy\cite{DFT84}.  On the other hand, for
nuclei near the proton drip line, the HF+BCS method is still
applicable in practice to obtain localized solutions because the
Coulomb barrier confines the wavefunctions of low-and-positive as well
as negative energy levels.


The contents of this paper are as follows.  In the
next section, we survey the features of the three-dimensional
Cartesian-mesh representation, discuss the choice of the interaction,
and explain the points newly developed in the present paper for
extensive calculations. Among these points are the determination of
the pairing force strengths and the acceleration of the convergence to
the solution.

In section \ref{s_mass}, we compare the atomic masses obtained from
our calculations with those in the latest experimental mass
table\cite{AW93}.  For this purpose, we consider a correction for
the error in the total binding energy due to the finite mesh size: A
greatly high precision is necessary in calculating the binding energy
compared with other quantities like moments.

Section~\ref{s_deform} treats the deformation.  We compare the
electric axial quadrupole moments with those deduced from the
experimental B(E2)$\uparrow$ data\cite{RMM87}.  We also discuss the
deformation parameters $a_{20}$, $a_{22}$, $a_{40}$, $a_{42}$, and
$a_{44}$, which we define for the HF+BCS solutions in terms of
multipole moments, and compare them with those of the FRDM.  The
difference of shapes between protons and neutrons as well as the
difference of the energies between the prolate and the oblate
solutions are also discussed.  We choose a shape coexisting nucleus
$^{80}_{40}$Zr$_{40}$ to illustrate the strong dependence on the force
parameters of the landscape of the potential energy curve along the
axial-quadrupole-deformation path.

The proton and the neutron skins are the subject of
section~\ref{s_skin}.  We discuss the thickness, the anisotropy, and
the relation with the halo radius.

In the last section, the contents of this paper is summarized and an
instruction to obtain our results electronically via computer network
is given.

\section{The set up of the calculations} \label{s_method}

\subsection{The basis} \label{s_basis}


The main feature of the HF+BCS code {\em EV8} used in this paper is
the three-dimensional Cartesian-mesh representation:
Each single-particle wavefunction $\psi(x,y,z)$ is defined in a
rectangular box
($-\frac{1}{2} L_x \le x \le \frac{1}{2} L_x$,
 $-\frac{1}{2} L_y \le y \le \frac{1}{2} L_y$,
 $-\frac{1}{2} L_z \le z \le \frac{1}{2} L_z$)
with its values $\psi_{ijk}$ at cubic mesh points, ($x_i,y_j,z_k$) =
$(i-\frac{1}{2}$, $j-\frac{1}{2}$, $k-\frac{1}{2})a$,
where $i$, $j$, and $k$ take on integers.  In this study, the
mesh size $a$ is set to 1 fm, while the size of the box is
$L_x=L_y=26$ fm, $L_z=28$ fm for $Z < 82$ and $L_x=L_y=28$ fm,
$L_z=30$ fm for $Z \ge 82$.  The nucleus is placed at the center of
the box.


We impose a symmetry with respect to reflections in $x$-$y$, $y$-$z$,
and $z$-$x$ planes (the point group D$_{\rm 2h}$). This symmetry
allows triaxial solutions, although all of our solutions have
eventually turned out axial and stable against $\gamma$-deformation.
On the other hand, the symmetry prohibits odd-multipole deformations,
which may not be negligible in some actinide nuclei.  According to the
calculations with the FRDM\cite{MNM94}, the nucleus
$^{222}_{\phantom{0}88}$Ra$_{134}$ has the largest octupole
deformation ($\beta_3$ = 0.15, the energy gain due to the octupole
deformation is $-1.4$ MeV), while except in the neighborhood of this
nucleus the octupole deformation occurs only in odd-$A$ and odd-odd
nuclei.  Incidentally, both in light\cite{KH95} and heavy
nuclei\cite{BKW91}, the octupole deformation is likely to be enhanced
by the procedure of the variation after parity projection.


One might wonder that a mesh size of 1 fm were too large to describe
the abrupt change of density at nuclear surface. It was demonstrated
in Ref.~\cite{BFH85}, however, that a mesh size $a$=1 fm can produce
enough accurate results for several spherical nuclei with mass below
$^{208}$Pb.  We did a similar test of accuracy for a deformed actinide
nucleus $^{240}_{\phantom{0}94}$Pu$_{146}$ and found that the relative
errors of the quadrupole moment and the total energy are 0.4\% and
0.5\%, respectively.  (The method of extrapolation to $a
\rightarrow 0$ is explained in section \ref{s_correction}.)  This
order of accuracy is higher than necessary for the quadrupole moment,
while it is not for the energy to make comparison with experiments.
Considering that the root-mean-square (r.m.s.)  deviation of the
atomic masses of recent mass formulae is $\sim$ 0.5 MeV, the desirable
precision is of the order of 0.1 MeV, which is only 0.005\% of the
total binding energy of $^{240}$Pu.  Therefore, the binding energy,
but not the other quantities, has to be corrected for the effect of
the finite mesh size, which is done in section~\ref{s_correction}.


The origin of this unexpectedly high accuracy with apparently coarse
meshes has been explained by Baye and Heenen\cite{BH86}.  The equation
to determine $\{ \psi_{ijk}
\}$ is usually derived through a discrete approximation to the
Schr{\"o}dinger equation. They presented an alternative point of view,
in which they introduced a set of orthogonal basis functions
$f_{ijk}(x,y,z)$ such that $\{ \psi_{ijk} \}$ are the coefficients to
expand $\psi(x,y,z)$ in this basis.  (In this point of view, the
equation for $\{ \psi_{ijk} \}$ is determined uniquely from the
variational principle.)  This basis can be unitary-transformed to
plane-wave basis with $\vert k_{\kappa} \vert < \pi / a$ ($\kappa$ =
$x$, $y$, $z$), which suggests that the atomic nucleus is a very
suitable physical object to apply the mesh representation because the
saturation property of nuclear matter guarantees the suppression of
large-momentum components in wavefunctions from the view point of the
Thomas-Fermi approximation.

To enjoy this high accuracy, the method to determine
$\{ \psi_{ijk} \}$ must be in accordance with the view point of Baye
and Heenen.  Exact variational treatment with the plane-wave basis
requires, however, long computation time and diminish the simplicity
of the Cartesian-mesh representation.  The code {\em EV8} is designed
to emulate the plane-wave expansion method, though it is based on the
discrete approximations, by choosing the appropriate orders of
approximation formulae for derivatives (the 7- and 9-point formulae
for the first and second derivatives, respectively) and integrals (the
mid-point formula).

It is worth mentioning that a new formulation of the mesh
representation in terms of collocation basis splines is being
developed recently by Chinn et al.\cite{CUV94}.

\subsection{The interaction} \label{s_interaction}


For the HF (mean-field) part of the interaction, the code uses the
Skyrme interaction\cite{Sk56,VB72}, which is a zero-range force with
the lowest order momentum dependences to emulate the finite-range
effects, a density dependence to reproduce the saturation of nuclear
matter density, and a spin-orbit coupling term.


The relation between the Skyrme-HF model and the relativistic
mean-field model\cite{Re89} has been discussed by many authors (see,
e.g., Ref.~\cite{SHT93}), which has invoked arguments on the density
dependence (the ratio of the isoscalar to the isovector density
dependences) of the spin-orbit term\cite{SLH94,DN94,PF94,RF94}.
However, as these arguments are not yet conclusive, we rather like to
use the old but well-examined standard form of the Skyrme force in
this paper.


Among the many parameter sets proposed for the Skyrme
force\cite{APD92,DFT84,BFG75,Koe76,KTO80,GS81,BQB82,WHW83,GPN92}, we
choose the SIII\cite{BFG75}.  Its validity has been examined in many
nuclear structure calculations.  In particular, it produces
single-particle spectra in good agreement with experiment.  It also
reproduces fairly well the $N-Z$ dependence of the binding
energy\cite{TBF93} compared with other widely-used parameter sets of
SGII\cite{GS81} and SkM$^{\ast}$\cite{BQB82}. Although its
incompressibility is said to be too large, it is not a serious
drawback because elaborate fittings of the parameters of the Skyrme
force\cite{RF94} and the FRDM\cite{MNM94} have shown that different
assumptions for the incompressibility lead to practically the same
quality of fittings to nuclear masses.


The force SkSC4\cite{APD92} was determined through the most extensive
fitting to nuclear mass data among the Skyrme forces. However, the
fitting was done in the ETFSI scheme, which produces unnegligibly
different energy from that of the HF method.  Because of this
disadvantage, we do not choose the force SkSC4 in this paper.

\subsection{The pairing} \label{s_pairing}


For the interaction in the pairing channel, we employ a seniority
force $V_{\rm pair}^{\tau}$, whose pair-scattering matrix elements are
defined as a constant multiplied by cutoff factors depending on the
single-particle energy $\epsilon_i$,
\begin{equation} \label{eq_vp}
  \langle i \bar{\imath} \vert V_{\rm pair}^{\tau} \vert
  j \bar{\jmath} \rangle = - G_{\tau} f_{\tau}(\epsilon_i)
  f_{\tau}(\epsilon_j),
\end{equation}
where $\tau$ signifies proton or neutron.
We assume the following form for the cutoff function,
\begin{equation} \label{eq_fe}
  f_{\tau}(\epsilon) = \left\lbrace 1 + \exp \frac{\epsilon -
  \epsilon_{\rm c}^{\tau}} {0.5 \; {\rm MeV}} \right\rbrace^{-1/2}
  \theta(e_{\rm c}^{\tau} - \epsilon),
\end{equation}
which contains two cutoff energies, $\epsilon_{\rm c}^{\tau}$ and
$e_{\rm c}^{\tau}$: The former is for a smooth cutoff necessary to
stabilize the iterative procedures to solve the HF equation, while the
latter is for a sharp cutoff preventing occupation of spatially
unlocalized single-particle states.  The cutoff energies take on the
following values,
\begin{equation} \label{eq_ec}
  \epsilon_{\rm c}^{\tau} = \lambda_{\rm HF}^{\tau}
  + 5 \; {\rm MeV}, \;\;\;\;
  e_{\rm c}^{\tau} = \epsilon_{\rm c}^{\tau}
  + 2.3 \; {\rm MeV},
\end{equation}
where $\lambda_{\rm HF}^{\tau}$ is the Fermi level defined as the
average of the highest occupied level and the lowest vacant level in
the HF (or normal) state.

In the BCS treatment of nuclei far from the $\beta$-stability line,
one has to take care so that the continuum states are not occupied,
which give rise to unphysical nucleon gas extending over the entire
box.  In our calculations, for neutrons, if the right-hand side of the
second of Eqs.~(\ref{eq_ec}) is positive, $e_{\rm c}^{\rm n}$ is
replaced by zero (for $Z < 82$) or $\epsilon_{\rm c}^{\rm n}$ is
replaced by $-2.3$ MeV (for $Z \ge 82$).
For protons, instead of lowering the sharp cutoff energy
$e_{\rm c}^{\rm p}$, we modify the proton potential outside the
Coulomb barrier (in the imaginary-time evolution operator, not in the
evaluation of the energy) so that the potential is higher than
$e_{\rm c}^{\rm p}$.  This prevents the tunneling through the barrier
and makes the proton single-particle wavefunctions whose energy is
within the pairing active interval ($\epsilon < e_{\rm c}^{\rm p}$)
spatially localized.


In order to treat a wide range of nuclei on a single footing, we need
a prescription to determine the strength $G_{\tau}$ for each nucleus.
For this purpose, we have developed a method based on the continuous
spectrum approximation.  We solve the following particle-number and
gap equations,
\begin{eqnarray} \label{eq_partnum}
N_{\tau} & = & \int_{-\infty}^{\infty} \left\lbrace
1 - \frac{\epsilon - \bar{\lambda}_{\tau}}
{\sqrt{(\epsilon-\bar{\lambda}_{\tau})^2 + f_{\tau}(\epsilon)^2
\bar{\Delta}_{\tau}^2}}
\right\rbrace \bar{D}(\epsilon) d \epsilon, \\
\bar{\Delta}_{\tau} & = & \frac{G_{\tau} \bar{\Delta}_{\tau}}{2}
\int_{-\infty}^{\infty} \frac{f_{\tau}(\epsilon)^2}
{\sqrt{(\epsilon-\bar{\lambda_{\tau}})^2 + f_{\tau}(\epsilon)^2
\bar{\Delta}_{\tau}^2}} \bar{D}(\epsilon) d \epsilon, \label{eq_gap}
\end{eqnarray}
which use the semiclassical single-particle level density
$\bar{D}(\epsilon)$ (obtained in the Thomas-Fermi approximation)
instead of the discrete spectrum of the single-particle HF
hamiltonian. One notes that $\bar{D}(\epsilon)$ does not have the
fluctuation due to the shell effects.

Eq.~(\ref{eq_partnum}) determines the Fermi level
$\bar{\lambda}_{\tau}$, while Eq.~(\ref{eq_gap}) tells the force
strength $G_{\tau}$ when it is combined with the empirical formula for
the pairing gap $\bar{\Delta}_{\tau}$ = $12 A^{-1/2}$ MeV.  For light
nuclei, the resulting $G_{\tau}$ becomes apparently too strong. We
replace $G_{\tau}$ with 0.6 MeV when it exceeds 0.6 MeV.  The
results are discussed in section \ref{s_gap}.

\subsection{The method of solution} \label{s_solution}


Because the single-particle basis of the mesh representation is huge,
it takes unmanageably long computation time to solve the HF+BCS
equation with the usual method of iterative diagonalization of the
single-particle HF hamiltonian $h_{\rm HF}$.  Instead, the code
{\em EV8} employs a much more efficient method of imaginary-time
evolution\cite{DFK80}, in which the evolution operator for a small
time interval $\Delta t$ is repeatedly operated on each
single-particle wavefunction to decrease its expectation value of the
energy. After each evolution, the single-particle wavefunctions are
orthogonalized in the Gram-Schmidt method from low to high energy
states.  As the initial wavefunctions, we utilize either the
eigenstates of the Nilsson model or the solutions for the neighboring
nuclei.


The imaginary-time evolution operator
          $\exp ( - \hbar^{-1} h_{\rm HF} \Delta t )$
is expanded to the first order in $\Delta t$ as
          $1 - \hbar^{-1} h_{\rm HF} \Delta t$.
To this order, the imaginary-time evolution method is
equivalent to the gradient iteration method\cite{Re91}.  Consequently,
to obtain the minimum energy state rather than the maximum one, it
must hold that
          $1 - \hbar^{-1} \epsilon_{\rm max} \Delta t > -1$,
where $\epsilon_{\rm max}$ is the maximum single-particle eigenenergy,
which is approximately the free kinetic energy for
          $k$ = $3^{1/2} \pi a^{-1}$.
With $a$=1 fm, it follows $\Delta t < 2.1 \times 10^{-24}$ sec (The
actual upper bound for $\Delta t$ is larger than this estimation
because the discrete approximation underestimates the kinetic energy
for very high momentum states).
We use $\Delta t=1.5 \times 10^{-24}$ sec in this paper.


We regard that the wavefunction is converged to a HF+BCS solution when
the following four criteria are satisfied.

i) The energy spreading of single-particle states are smaller than 0.1
MeV.  Practically, we require that the third largest value of
\begin{equation} \label{eq_dei}
  \left( \Delta \epsilon_i \right)^2 \equiv
  \left( \langle i \vert h_{\rm HF}^2 \vert i \rangle -
  \langle i \vert h_{\rm HF} \vert i \rangle^2 \right)
  \cdot \min \left( 2 v_i^2, 1 \right)
\end{equation}
should be less than (0.1 MeV)$^2$, where $\vert i \rangle$ represents
the $i$th single-particle state and $v_i^2$ its BCS occupation
probability.

ii) During the last 75 steps of evolution, the variation in the total
energy is $<$ 1.5 keV for $Z < 82$ and $<$ 5 keV for $Z \ge 82$.

iii) During the same period, the variation in
$\langle x^2 \rangle_{\tau}$,
$\langle y^2 \rangle_{\tau}$, and
$\langle z^2 \rangle_{\tau}$
are $<$ 0.5 fm$^2$ for $Z < 82$ and
$<$ 0.005 $\times$ $\frac{1}{5}r_0^2 N_{\tau} A^{2/3}$ for $Z \ge 82$,
where $\tau$ stands for all the protons or all the neutrons
and $r_0 \equiv 1.2$ fm.

iv) The axial quadrupole deformation parameter $\delta$ is close to
the convergent value.  We adopt a definition
$\delta$ $\equiv$ 3 $\langle Q_z\rangle$ /
                  4$\langle r^2 \rangle$\cite{BM75},
where $Q_z$ $\equiv$ $2 z^2 - x^2 - y^2$.
The convergent value, $\delta_{\rm pred}$, together with its
uncertainty, $\Delta \delta_{\rm pred}$, are predicted using the
values of $\delta$ during the last 200
steps\footnote{ In this paper,
$\delta_{\rm pred}$ is determined by an equation $\dot{\delta}$=0,
where $\dot{\delta}$ is the time derivative of $\delta$ and expressed
as a quadratic function in $\delta$ whose coefficients are determined
from the least-square fitting to the last $n$ steps.  Changing $n$
between 200 and 150 produces a set of values $\{ \delta_{\rm pred}
\}$.  The average of them is adopted as $\delta_{\rm pred}$, while a
half of the difference between the maximum and the minimum of them is
used as the size of the error $\Delta \delta_{\rm pred}$.
}.
The condition is expressed as $\vert \delta$ $-$ $\delta_{\rm pred}
\vert$ + $\Delta \delta_{\rm pred}$ $<$ $0.004$.  For nuclei in the
shape-transitional region this criterion is the last one to be
satisfied among the four, because the potential energy surface (PES)
is very flat versus $\delta$.


In order to accelerate the convergence when the wavefunction looks
slowly converging to a state with deformation $\delta_{\rm pred}$, we
exert an external mass quadrupole potential during several tens of
evolution steps so that the wavefunction quickly acquires the
predicted deformation.  Then, we switch off the external potential and
continue the free imaginary-time evolution.  With this acceleration
method, the necessary number of evolution steps to achieve convergence
can be decreased by a factor larger than two.  The reason why this
method works is that the quadrupole deformation is almost always the
softest mode in atomic nuclei (under the D$_{\rm 2h}$ symmetry).  On
the other hand, in generic multi-dimensional minimization problems,
the principal difficulty is in finding the direction of the softest
mode\cite{PFT89}.


In the top portion of Fig.~\ref{f_conv} we show histories of the
imaginary-time evolution of the deformation parameter $\delta$ with
(solid curve) and without (dot curve) the acceleration method for the
prolate solution of $^{156}_{\phantom{0}68}$Er$_{88}$.  The sharp fall
of the solid curve is due to the external quadrupole potential exerted
between the time steps 213 and 291. One can see that the convergence
can be achieved within much shorter time than without the acceleration
method.

In the bottom-left and bottom-right portions of the figure, the time
evolution of the maximum spreading in the single-particle levels,
Eq.~(\ref{eq_dei}), and the total energy (measured from the convergent
value) are displayed, respectively, for the same process as in the top
portion.  One can see that the convergence of these quantities are
also advanced by using the external potential.


\begin{center}
\framebox[4cm]{Figure \ref{f_conv}}
\end{center}


All the solutions of our extensive HF+BCS calculation have been found
axially symmetric. The stability of these solutions against triaxial
deformation can be known from the time evolution of the (very small)
triaxiality parameter when the external potential is not effective,
because it should grow exponentially if the axial solution is a saddle
point rather than a minimum.  It has turned out that none of the axial
solutions exhibit clearly such an exponential growth of the
triaxiality.  It is possible, however, that triaxial minima exist
which are separated from the axial path by a potential barrier.

Bonche et al.\ found triaxial ground-state solutions for $^{84}$Zr and
$^{94}$Zr using the same Skyrme interaction and the same computer code
as we used.  However, these triaxial minima look so shallow that they
can easily be moved to axially symmetric shapes by, e.g., making the
pairing correlation slightly stronger.


The treatment of the Coulomb interaction is in line with the
appendix~C of Ref.~\cite{FKW78}: The exchange part of the interaction
is approximated by a local potential in terms of the Slater
approximation.  The direct part is implemented in terms of a Coulomb
potential obtained by solving the Poisson equation using a three-point
approximation to the second derivative. The boundary conditions are
determined by, first, dividing the nucleus into two fragments in one
of the $x$-$y$, $y$-$z$, and $z$-$x$ planes (we select the plane which
gives the largest distance between the centers of mass of the two
fragments) and, second, expanding the Coulomb potential originating in
each fragment in terms of multipole moments of protons through order
three.

Throughout this paper, we treat a nucleon as a point particle,
neglecting its extension.  The effect of the spurious center of mass
motion is corrected for by reducing the bare nucleon mass by a factor
$1-A^{-1}$.  The explicit form of the hamiltonian density is found in
Ref.~\cite{BFH85}.

\section{The nuclear masses} \label{s_mass}


In the manner described in section~\ref{s_method}, we have solved the
HF+BCS equation for even-even nuclei with $2 \le Z \le 114$.  For each
isotope chain, the calculation extends from outside the proton drip
line by several neutrons to beyond the experimental frontier in the
neutron-rich side by a few neutrons. For $Z > 100$, the calculation
extends to $N$ $\le$ $2 Z - 42$.  Spatially localized solutions have
been obtained for 1029 nuclei.  Although some of these solutions have
negative two-proton separation energies, their densities are localized
owing to the Coulomb barrier.  The second local minima have been found
for 758 nuclei.


We determine the ground-state solution of each nucleus by, first,
searching the spherical, a prolate, and an oblate solutions and,
second, comparing the energies of thus obtained solutions.  Our
strategy to search for these three solutions for each nucleus is as
follows. The spherical solution is obtained by constraining the mass
quadrupole moments to be zero.  The prolate (oblate) solution is
searched in two steps.  First, we exert an external potential
proportional to $Q_z$ on the initial wavefunction until its quadrupole
deformation parameter satisfies $\delta > 0.1$ ($< -0.1$).  Second, we
switch off the external potential, let the wavefunction evolve by
itself (or with the acceleration method described in
section~\ref{s_solution}), and see if it converges to a deformed local
minimum.  If the nuclear shape becomes very close to the sphericity in
the course of evolution, i.e. $\delta < 0.02$ ($> -0.02$), we conclude
that the normal-deformation prolate (oblate) solution does not exist
in this nucleus.


For some nuclei with $28 < Z, N < 50$, the FRDM\cite{MNM94} predicts
very large deformations $\delta \sim 0.4$.  In order not to miss such
large-deformation solutions, we have done additional searches for all
the nuclei in this region, in which we continue to exert the
quadrupole potential until $\delta$ becomes $>0.4$ ($< -0.3$) before
starting the free evolution for the prolate (oblate) solution.  These
additional searches indeed produced large-deformation solutions.
However, none of them are the ground states unlike in the results of
the FRDM.


In shape-transitional nuclei, the PES often has more than three
normal-deformation minima.  However, they are usually very shallow and
it is doubtful that each of them corresponds to a distinct eigenstate
notwithstanding the quantum fluctuation in shape.  Therefore, we do
not manage to find out all of these shallow minima.

\subsection{Correction for the finite mesh size} \label{s_correction}


As we have discussed in section~\ref{s_basis}, it is necessary to
correct the total energy for the inaccuracy due to the finite mesh
size because its relative error has to be by far smaller than that of
other quantities for the sake of comparison with experimental data.


To evaluate the size of the error, one needs the HF+BCS solution for
vanishingly small mesh size.  It can be obtained without difficulty
concerning spherical solutions, because even personal computers can
execute spherical HF+BCS codes with very small radial-grid spacing.
Therefore, we have chosen to compare the total energies between the
solutions of a spherical HF+BCS code\footnote{
The spherical code has been modified such that the following points
are the same as in the Cartesian code: the values of physical
constants, the treatment of the pairing correlations, and the
modification of the proton potential outside the Coulomb barrier. The
volume of the spherical cavity was set to the same size as that of the
rectangular box used in the Cartesian code.  The radial-grid spacing
was set at 0.1 fm.}
(the {\em SKHAFO} taken from Ref.~\cite{Re91})
and the Cartesian-mesh code {\em EV8} with constraints of vanishing
quadrupole moments.
In the following, we designate the total energy obtained with the
spherical code as $E_0$ while that from the Cartesian-mesh code simply
as $E$.  Our aim is to construct a formula for the energy correction
so that the corrected energy $E_{\rm c} = E +$(correction) has a much
smaller r.m.s.\ deviation from $E_0$ than $E$ has.

In the least-square fitting to determine the parameters of the
formula, we have used 1005 nuclei (among the 1029 nuclei) whose paring
gaps (both for proton and neutron) coincide within 0.1 MeV between the
results of the two codes.  With the simplest fitting function of
$E_{\rm c}$ = $E$ + $c_1 A$, the r.m.s. difference between $E_{\rm c}$
and $E_0$ can be decreased from 6.7 MeV to 0.35 MeV.
By adding temrs up to the second order in $N$ and $Z$,
\begin{equation} \label{eq_fit5}
  E_{\rm c}= E + c_1 A + c_2 (N-Z) + c_3 A^2 + c_4 A(N-Z)
        + c_5 (N-Z)^2,
\end{equation}
the r.m.s.\ value of $E_{\rm c} - E_0$ is decreased to 114 keV with
$c_1=-40.2$, $c_2=-20.6$, $c_3= 0.033$, $c_4=-0.081$, and
$c_5=-0.080$ (keV).  Since this size of error is much smaller than
the typical precision of the mass formulae in the market place
($\sim 0.5$ MeV)\cite{MNM94,APD92,TUY88}, we have decided to
adopt the above formula\footnote{
Inclusion of higher order terms of $N$ and $Z$ to the correction
formula does not substantially decrease the r.m.s.\ error.  With terms
of degrees from zero to three (10 terms in total), the r.m.s.\ error
is 107 keV. Addition of up to sixth-order terms (28 terms) leads to an
r.m.s.\ error of 93 keV.  On the other hand, addition of only one more
term $c_6 E_{\rm TD} A^{1/3}$ decreases the r.m.s.\ error to 86 keV,
where $E_{\rm TD}$ is the space integral of
$B_5 \rho         \triangle \rho        $ +
$B_6 \rho_{\rm n} \triangle \rho_{\rm n}$ +
$B_6 \rho_{\rm p} \triangle \rho_{\rm p}$
(see Ref.~\cite{BFH85} for the definitions of $B_5$ and $B_6$).
Further addition of a term
$c_7 (\Delta_{\rm n} + \Delta_{\rm p}) A$
reduces the error to 77 keV, where
$\Delta_{\rm n}$ and $\Delta_{\rm p}$
are the pairing gaps.
}.

We have tested the accuracy of the correction formula (\ref{eq_fit5})
for deformed solutions by comparing $E_{\rm c}$ with an energy
extrapolated to $a \rightarrow 0$.  In the extrapolation, first, we
calculate the total energy for
seven values of $a$ ranging from 1 fm to 0.56 fm (six values ranging
from 1 fm to 0.6 fm for $^{240}$Pu), while keeping the box size
constant.  Second, we fit a polynomial
$E(a) = E_{\rm ext} + b_1 a^2 + b_2 a^6$
to the seven or six sets of values ($a$, $E$) using
$E_{\rm ext}$, $b_1$, and $b_2$ as the fitting parameters.
This form of the fitting function has been chosen on the following
grounds: At $a \sim$ 1 fm, the error is dominated by a term of order
$a^6$, which originates in the seven-point approximation to the first
derivatives.  At $a \sim$ 0.5 fm, the contribution from lower-order
terms becomes comparable to that of the $a^6$ term. These terms seem
to arise principally from the error in the Coulomb energy, whose
leading order term is $a^2$.  The ambiguity in the extrapolated energy
$E_{\rm ext}$ is roughly estimated to be $\sim 0.2$ MeV.

In Table~\ref{t_becorr}, the difference between the extrapolated
energy to $a \rightarrow 0$ and the energy calculated with $a$=1 fm is
shown for oblate, spherical (obtained with constraints of vanishing
mass quadrupole moments), and prolate solutions of five nuclei.  The
fifth column shows the energy correction given by
formula~(\ref{eq_fit5}).  For these 15 solutions, the mean and the
r.m.s.\ values of the difference between the correction formula and
the extrapolations are 0.0 MeV and 0.21 MeV, respectively, which are
smaller than or of the same size as the accuracy of the extrapolation.
This agreement shows the applicability of the correction
formula~(\ref{eq_fit5}) to deformed solutions as well as to spherical
ones.  It also confirms that the spherical code and the Cartesian code
with the spherical constraint are indeed equivalent to each other.


\begin{center}
\framebox[4cm]{Table \ref{t_becorr}}
\end{center}

\subsection{Comparison of the masses} \label{ss_mass}


Fig.~\ref{f_masx} presents the corrected total energies $E_{\rm c}$ of
1029 even-even nuclei calculated with the HF+BCS method with the
Skyrme SIII force\footnote{
%
%
The nuclear binding energy corresponds to $-E_{\rm c}$, the nuclear
mass to $E_{\rm c} + Z m_{\rm p} + N m_{\rm n}$, and the atomic mass
to $Z m_{\rm e} - B_{\rm e}(Z)$ added by the nuclear mass.
The values of $m_{\rm p}$, $m_{\rm n}$, and $m_{\rm e}$ are taken from
Ref.~\cite{AW93}. We use $B_{\rm e}(Z)$=14.33 $Z^{2.39}$ eV. The
definitions of energy-related quantities like $S_{\rm 2p}$ and
$S_{\rm 2n}$ are given in Ref.~\cite{AW93b}.
}.
For graphical reason, the smooth (i.e., macroscopic) part
$E_{\rm macro}$ has been subtracted, which is defined by fitting
functions of the Bethe-Weizs{\"a}cker type,
\begin{equation} \label{eq_bw}
  E_{\rm macro} = a_{\rm V} A + a_{\rm S} A^{2/3}
                + a_{\rm I} (N-Z)^2 A^{-1} + a_{\rm C} Z^2 A^{-1/3},
\end{equation}
to $E_{\rm c}$ separately for $A < 50$ and $A \ge 50$, varying
$a_{\rm V}$, $a_{\rm S}$, $a_{\rm I}$, and $a_{\rm C}$ as free fitting
parameters.
The solid (open) circles designate that the nucleus is more (less)
bound than the macroscopic trend $E_{\rm macro}$, while the diameter
of each circle is proportional to
$\vert E_{\rm c} - E_{\rm macro} \vert$.

We also show the two-proton (two-neutron) drip lines for the HF+BCS
with SIII and for the FRDM\cite{MNM94}.  The two-proton (two-neutron)
drip line lies between two even-even nuclei whose two-proton
(two-neutron) separation energies $S_{\rm 2p}$ ($S_{\rm 2n}$) have
different signs.  As for the two-neutron drip line for the SIII
force, we use the macroscopic energy (Eq.~(\ref{eq_bw})) fitted to
the SIII results
($a_{\rm V}$ =$-14.702$ MeV,
 $a_{\rm S}$ =  14.05   MeV,
 $a_{\rm I}$ =  21.47   MeV,
 $a_{\rm C}$ =   0.6554 MeV)
because our calculation does not extend to the neutron drip line.


One can see regions of enhanced stability around double-magic nuclei
with $(N,Z)$ = $(50,50)$, $(82,50)$, and $(126,82)$.  Another
double-magic nucleus $(82,82)$ is outside the two-proton drip line.
The super-heavy double-magic nucleus $(184,114)$ does not look like a
local minimum of nuclear mass in this result.


The two-proton drip lines of the HF+BCS with SIII (solid line) and the
FRDM (dash line) are overlapping in most places. The distance
between them is $\Delta Z=4$ for $N=40$, $\Delta N=4$ for $Z=$42 and
78, and $\Delta Z \le 2$ and $\Delta N \le 2$ for the other isotope
and isotone chains.

The two-neutron drip lines of the two theoretical approaches are also
close to each other.  The difference looks of the same size as
that between the FRDM and the TUYY mass formula\cite{TUY88}, both of
which are models whose parameters were determined through extensive
fittings to the mass data. This fact indicates the quantitative
appropriateness of the macroscopic isospin dependence of the SIII
force.  Indeed, it is what we expected in choosing the SIII force for
our first extensive HF+BCS calculation.


\begin{center}
\framebox[4cm]{Figure \ref{f_masx}}
\end{center}


The r.m.s.\ deviation of the calculated ground-state masses from the
experimental ones of 480 even-even nuclei (the best recommended values
of Ref.~\cite{AW93} excluding those estimated from systematic trends)
is 2.2 MeV.  Note that the inaccuracy of calculations due to the
finite mesh size remaining after the correction is by far smaller than
this deviation.


The difference for each nucleus is shown in Fig.~\ref{f_masxdf}.  The
solid (open) circles are put when the calculated masses are smaller
(larger) than the experimental ones, while the diameter of the circles
is proportional to the magnitude of the difference.  One can see that
the calculated masses tend to be overbinding for $Z$=8 and 20
isotopes, and $N$=50, 82, and 126 isotones.  Unlike spherical
nuclei including these semi-magic isotopes and isotones,
deformed nuclei have positive errors, which are $\sim$ 3 MeV
rather independently of the size of deformation.


\begin{center}
\framebox[4cm]{Figure \ref{f_masxdf}}
\end{center}


In Table~\ref{t_massdev}, we show the r.m.s.\ differences of the
masses of even-even nuclei between theoretical models as well as
between the experiments and the models.  In the table, {\bf AW'93}
represents the experimental atomic mass table by Audi and
Wapstra\cite{AW93}, {\bf TUYY} the mass formula of Tachibana et
al.~\cite{TUY88}, {\bf FRDM} the finite-range droplet
model\cite{MNM94}, {\bf ETFSI} the extended Thomas-Fermi Strutinsky
integral method with the SkSC4 force\cite{APD92}, {\bf EV8C} the
HF+BCS results using the Skyrme SIII force with the correction
(Eq.~(\ref{eq_fit5})), and {\bf macro} the Bethe-Weizs{\"a}cker type
function fitted to AW'93 ($a_{\rm V}$ =$-15.280$ MeV, $a_{\rm S}$ =
16.01 MeV, $a_{\rm I}$ = 22.33 MeV, $a_{\rm C}$ = 0.6896 MeV).  In
parentheses are the number of nuclei to calculate the difference.  The
r.m.s.\ deviation from AW'93 is 2.2 MeV for EV8C, which is 3-4 times
as large as that of 0.52 MeV for TUYY, 0.68 MeV for FRDM, and 0.74 MeV
for ETFSI.  It should be noticed, however, that the parameters of
FRDM, TUYY, and ETFSI were fitted to all the available recent
experimental mass data while the parameters of the SIII force were
determined by fitting to the masses and charge radii of only seven
spherical nuclei.  In addition, the number of the fitting parameters
is 275 in TUYY, 19 in FRDM, and 8 in ETFSI, while it is only 6 in the
SIII force.


\begin{center}
\framebox[4cm]{Table \ref{t_massdev}}
\end{center}


As a further investigation of the macroscopic properties of the mass,
we examine the possibility to decrease the r.m.s.\ deviation by
improving the macroscopic part of the mass models.  Namely, for each
combination of the nuclear mass models and the experimental data, we
add the Bethe-Weizs{\"a}cker type function (Eq.~(\ref{eq_bw})) to one
of them and determine the four coefficients to minimize the
r.m.s.\ difference between them.  The resulting r.m.s.\ differences
are tabulated in Table~\ref{t_improv}.  This simple correction method
can decrease the r.m.s.\ error of EV8C from experiments by 27\%. As
for the other models, however, the improvements are marginal.  On the
other hand, the differences between the models are greatly decreased
because they come predominantly from nuclei near the neutron drip
line.  This fact suggests that substantial improvements of the
macroscopic part of the nuclear mass models are possible only if new
experimental mass data of neutron-rich nuclei are provided.


\begin{center}
\framebox[4cm]{Table \ref{t_improv}}
\end{center}

\subsection{Pairing gaps} \label{s_gap}

Let us compare the experimental and the calculated pairing gaps.  The
former are calculated by applying the 5-point formula\cite{MN88} to
the experimental masses\cite{AW93} of nuclei except those at
major-shell closures (of protons for the proton gap, of neutrons for
the neutron gap), while the latter are obtained directly from the BCS
part of the HF+BCS scheme.

When the gaps are plotted versus the mass number $A$, the experimental
ones fall close to the empirical formula $\Delta$ = 12 $A^{-1/2}$ MeV.
The calculated gaps also fall near the empirical curve for heavy
nuclei but they are raised for $50 < A < 100$. In lighter nuclei ($A <
50$), the calculated gaps are located below the curve, which is due to
our setting the maximum pairing force strength to 0.6 MeV.  A likely
origin of the overestimation for $50 < A < 100$ nuclei is that the
pairing active space above the Fermi level (see Eqs.~(\ref{eq_ec})) is
smaller than the wavelength of the shell oscillation ($\hbar
\omega_{\rm osc}$ $\sim$ $41 A^{-1/3}$ MeV) for small $A$; such a
situation does not fit to the continuous spectrum approximation.  In
future calculations for $A<100$, it is desirable to improve the method
to determine the pairing force strengths.

\section{Deformation} \label{s_deform}


Since we impose the reflection symmetries in $x$-$y$, $y$-$z$,
$z$-$x$-planes (the D$_{\rm 2h}$ symmetry) on our HF+BCS solutions, as
explained in section~\ref{s_basis}, our solutions have only multipole
moments with even angular momentum $l$ and even magnetic quantum
number $m$.

\subsection{quadrupole moments} \label{s_q}


In Fig.~\ref{f_qq}, we compare the magnitudes of the intrinsic
electric quadrupole moments between our HF+BCS solutions and the
experimental values deduced from B(E2)$\uparrow$, i.e., the reduced
electric quadrupole transition probability from the ground state to
the first excited $2^{+}$ state \cite{RMM87}.  This moment is defined
for the HF+BCS solutions as
\begin{equation} \label{eq_q0}
Q_0 = \sum_{\rm protons} \left\langle 2 z^2 - x^2 - y^2 \right\rangle,
\end{equation}
where $z$-axis is the symmetry axis (All of our solutions have
practically axial shapes, as described in section \ref{s_defparm}).
The deduction of $\vert Q_0 \vert$ from the B(E2)$\uparrow$ is based
on the rigid rotor model \cite{RMM87,BM75}.

For nuclei with $Q_0 > 3.5$~b, the agreement with experiment is
excellent.  The even-even nuclei having the largest intrinsic
quadrupole moment is $^{252}_{\phantom{0}98}$Cf$_{154}$ (indicated by
letter {\bf A} in the figure) in the experimental table\cite{RMM87},
while it is $^{244}_{106}$Rf$_{138}$ among the 1029 HF+BCS
solutions ($Q_0$=16.6~b).

The largest discrepancy is found in $^{222}_{\phantom{0}90}$Th$_{132}$
(indicated by {\bf D}), whose experimental $Q_0$ is 5.5~b while the HF
solution is spherical.  Two nuclei located at isolated points are
$^{176}_{\phantom{0}78}$Pt$_{98}$ (indicated by {\bf B}) and
$^{222}_{\phantom{0}88}$Ra$_{134}$ (indicated by {\bf C}).

For nuclei with smaller $Q_0$, however, many nuclei falls not in the
diagonal line but in a horizontal line, which means that the HF
solution has a spherical shape when the experimental B(E2)$\uparrow$
is not necessarily very small.  This may be explained by attributing
the enhanced B(E2)$\uparrow$ not only to static deformations but also
to the collective shape oscillation around the spherical equilibrium.
It may also be related to the complicated landscapes of the potential
energy curves of nuclei with $A$=50-100 (see section~\ref{s_zr}).


\begin{center}
\framebox[4cm]{Figure \ref{f_qq}}
\end{center}

\subsection{Deformation parameters} \label{s_defparm}


For certain purposes, the deformation parameters are more useful than
the electric multipole moments, although the former are
model-dependent while the latter are directly related to the
experimental observables.  In a widely-used method of the Strutinsky
shell correction, the deformation parameters are built in the theory
in order to specify the single-particle potential.  On the other hand,
for mean-field solutions, one has to define the deformation parameters
from the density distributions of nucleons.

In this paper, we define the deformation parameters as those of a
sharp-surface uniform-density liquid drop which has the same mass
moments as the HF+BCS solution has.  The mass density of the liquid
drop is expressed as
\begin{eqnarray} \label{eq_rhold}
  \rho(\bm{r}) & = & \rho_0 \; \theta ( R(\hat{\bm{r}})-|\bm{r} |),\\
  R(\hat{\bm{r}}) & = & R_0 \; \Bigl( 1 + \sum_{l,m}
  a_{lm} Y_{lm} (\hat{\bm{r}}) \Bigr). \label{eq_radld}
\end{eqnarray}
The necessary and sufficient conditions on $a_{lm}$ to fulfill the
reality of $R(\hat{\bm{r}})$ and the D$_{\rm 2h}$ symmetry are that
$l$ and $m$ are even numbers and $a_{lm}=a_{lm}^{\ast}=a_{l-m}$.  We
set $a_{lm}=0$ for $l \ge 6$ and determine the remaining seven
parameters $\rho_0$, $R_0$, $a_{20}$, $a_{22}$, $a_{40}$, $a_{42}$,
and $a_{44}$ such that the liquid drop has the same particle number,
mean-square mass radius, and mass quadrupole ($r^2 Y_{2m}$) and
hexadecapole ($r^4 Y_{4m}$) moments\footnote{
One might wonder that the radial dependence of $r^4$ so strongly
emphasizes the contributions from the peripheral regions that the
deformation is sensibly affected by the shape of the box. To see the
size of such erroneous effects we have done calculations by changing
the radial dependence of the hexadecapole moment from $r^4$ to $r^2$
in the determination of the deformation parameters. The resulting
values of $a_{lm}$ are only marginally changed except in a few nuclei
outside the proton drip line. Incidentally, the FORTRAN source code to
determine the seven liquid-drop parameters is obtainable via computer
network.  See an explanation at the end of this paper.
}
as the HF+BCS solution has.

The resulting values of $R_0$ and $a_{20}$ coincide very well with
$(\frac{5}{3})^{1/2} r_{\rm rms}$ and $\delta$, respectively, where
$r_{\rm rms}$ is the r.m.s.\ radius and $\delta$ is the deformation
parameter defined in section~\ref{s_solution}.  For 5361 samples
(mass, proton, and neutron moments of 1029 ground and 758
first-excited solutions), the maximum and the r.m.s.\ deviations of
$R_0$ from $(\frac{5}{3})^{1/2} r_{\rm rms}$ are 0.3 fm and 0.06 fm,
respectively. Those of $a_{20}$ from a fitted function
$(\frac{16}{45}\pi)^{1/2} \delta$ $-0.47 \delta^2$ $+0.78 \delta^3$ is
0.05 and 0.007, respectively.


The axial quadrupole deformation parameter $a_{20}$ is shown in
Fig.~\ref{f_a20} for the ground-state solutions of the HF+BCS equation
with the SIII force.  The open (solid) circles designate prolate
(oblate) nuclei, while the diameter of the circles is proportional to
the magnitude of the deformation parameter.  The two-proton and
two-neutron drip lines from our HF+BCS calculations (same as in
Fig.~\ref{f_masx}) are drawn for the sake of convenience.

One can see that nuclei at major-shell closures are spherical except
for some $N$=28 isotones, while the deformation develops between the
major-shell closures.  Such pattern of the development of $a_{20}$
looks regular for $A > 100$, while it is not so regular for $A < 100$,
i.e., the change of deformation along isotope or isotone chains is not
always smooth. In addition, oblate ground states are embedded here and
there in light-mass region while for heavier-mass nuclei they are
found only exceptionally in regions near major-shell closures.

The largest deviation from the experimental $a_{20}$ deduced from
B(E2)$\uparrow$ occurs in $^{12}$C.  For this nucleus the experimental
B(E2)$\uparrow$ is very large (corresponding to $\vert a_{20} \vert
$=0.59) and an oblate intrinsic deformation with a triangular
three-alpha-cluster configuration has been suggested.  Indeed,
calculations using the Nilsson model\cite{Vo74} and the Strutinsky
method\cite{LL75} give oblate ground states. On the other hand, the
HF+BCS calculation with the SIII force gives a potential energy curve
which has only one minimum at the spherical shape.  Other widely-used
Skyrme forces of the SkM$^{\ast}$ and the SGII also give the only
minimum at sphericity, while an old Skyrme force SII gives an oblate
minimum with $\delta=-0.27$\cite{Va73}.  However, one cannot deduce
the superiority of the SII force since the optimal shapes of light
nuclei are apt to be changed when effects beyond mean-field
approximations are taken into account.  For example, the parity
projection may be important\cite{KH95} because the triangular
three-alpha-cluster configuration violates the symmetry.

A comparison with the results of the FRDM\cite{MNM94} indicates a
systematic differences in $Z \sim N \sim 40$ region, where our
solutions tend to predict smaller deformations than those of the FRDM.
We discuss on this difference in section~\ref{s_zr}.
Another systematic difference occurs in a long and narrow region close
to the proton drip line with $94 \le Z \le 102$, where the FRDM
predicts oblate shapes while our calculations give small prolate
shapes, on which we do not discuss in this paper.


\begin{center}
\framebox[4cm]{Figure \ref{f_a20}}
\end{center}


Deformations with the magnetic quantum number $m$=2 (triaxial
deformations) are almost vanishing for all the nuclei we calculated:
The magnitudes of $a_{22}$ and $a_{42}$ are smaller than $10^{-4}$.
The deformation with $m$=4 ($a_{44}$) is larger than those with $m$=2.
Its typical size is of the order of $10^{-3}$ and the sign is mostly
negative.  However, these small non-axial deformations cannot affect
any observables in experimentally detectable ways.  Incidentally, it
is an interesting problem how $a_{44}$ develops as the angular
momentum increases\cite{HM94}.


The axial hexadecapole deformation parameter $a_{40}$ is shown in
Fig.~\ref{f_a40} in the similar manner as in Fig.~\ref{f_a20}.  This
parameter becomes sizable for $Z > 50$.  The sign of $a_{40}$ is
positive in the first half of the major shells and negative in the
second half. This behavior is in agreement with the results of the
FRDM\cite{MNM94} as well as with a naive expectation from the density
profile of pure-$j$ single-particle wavefunctions.  The largest value
of $\vert a_{40} \vert$ is about 0.1 both for positive and negative
signs.


\begin{center}
\framebox[4cm]{Figure \ref{f_a40}}
\end{center}


One of the advantages of mean-field methods over shell-correction
schemes is that the protons and the neutrons do not have to possess
the same radius and deformation.  Making use of this advantage, we
have calculated the liquid-drop shape parameters separately for
protons and neutrons for 1029 ground and 758 first-excited solutions.
As for $a_{20}$, the r.m.s.\ difference between protons and neutrons
is 0.012, while the maximum difference occurs in the ground state of
$^{14}_{\phantom{0}4}$Be$_{10}$ ($a_{20}^{\rm pro}=0.52$, $a_{20}^{\rm
neu}=0.36$).  The absolute value of the difference is smaller than
0.05 for 98.9 \% of the solutions.  For $a_{40}$, the r.m.s.\
difference is 0.0063, while the maximum difference is found in
$^{16}_{\phantom{0}4}$Be$_{12}$ ($a_{40}^{\rm pro}=-0.03$,
$a_{40}^{\rm neu}=-0.13$).  Concerning $R_0$, the r.m.s.\ difference
is 0.18 fm. The maximum difference excluding nuclei outside the drip
lines occurs again in $^{16}$Be ($R_0^{\rm pro}=3.0$ fm, $R_0^{\rm
neu}=3.9$ fm).  To summarize, the shapes of proton and neutron density
distributions are not remarkably different from each other in any of
our solutions.  As for the radius, we also study the difference at
much lower densities, i.e., skins and halos, in section~\ref{s_skin}.

\subsection{$28 < Z, N < 50$ nuclei} \label{s_zr}


Our calculations tend to predict smaller deformations than the
FRDM\cite{MNM94} for nuclei in $28 < Z, N < 50$.  These discrepancies
in deformation often occur in a manner that our model gives a
spherical shape while the FRDM predicts $\delta \sim 0.4$.  These are
originated in shape coexistence, i.e., the potential energy curve has
more than one minimum which are energetically competing within $\sim
1$ MeV.


As an example to illustrate such a subtle situation, we take up
$^{80}_{40}$Zr$_{40}$, for which the FRDM predicts a large prolate
deformation $a_{20}$=0.43 while our calculation gives a moderate
oblate deformation.  In Fig.~\ref{f_zrpes} the solid curve represents
the potential energy curve calculated with the SIII force.  It has as
many as three minima, i.e., an oblate one at $\delta=-0.18$, almost
spherical one, and a prolate one at $\delta=0.41$.  Because the
energies of these three minima are so close to each other (they are
within 0.6 MeV), the order of the energies can be altered easily by
changing the parameters of the interaction.  With our chosen
parameters, the oblate minimum has the lowest energy (solid curve). If
one decreases the pairing gap by 25\% (i.e., the average pairing gap
$\bar{\Delta}_{\tau}$ with which to determine the pairing force
strength is changed from 12 $A^{-1/2}$ MeV to 9 $A^{-1/2}$ MeV) the
prolate minimum becomes the ground state (dot curve).  Instead, by
changing the Skyrme force parameter set to the SkM$^{\ast}$, while
using the standard pairing force strength, one can make the prolate
minimum the ground state (dash curve).  On the contrary, the SGII
force deepens the spherical minimum (dot-dash curve).  The shapes of
nuclei in this mass region have been studied in many
papers\cite{GDG89,ZZ91,MGS92,KFP93}.


\begin{center}
\framebox[4cm]{Figure \ref{f_zrpes}}
\end{center}

\subsection{Oblate solutions} \label{oblate}

In Fig.~\ref{f_oblpro}, the energy differences between the oblate and
the prolate solutions are plotted versus the neutron number.  We put
circles for those nuclei which have both the prolate and the oblate
local minima.  The circles belonging to the same isotope chain are
connected by a line to guide the eyes.

The energy difference is small near major-shell closures but large in
the middle of the major shells. Apart from this shell fluctuation, one
can see some overall trends. For each of the three major shells of
neutrons divided by $N$=50, 82, 126, and 184, the largest energy
difference is
4.70 MeV ($^{124}_{\phantom{0}62}$Sm$_{62}$, $\delta$=$-0.25, 0.37$),
6.58 MeV ($^{164}_{\phantom{0}64}$Gd$_{100}$,$\delta$=$-0.22, 0.31$),
and 11.05 MeV ($^{254}_{102}$No$_{152}$, $\delta$=$-0.17, 0.26$),
respectively; as the nucleus becomes heavier, the energy difference
increases while the size of deformation decreases.

For nuclei with $N < 50$, the changes along isotope chains are not so
regular as in heavier nuclei and the oblate solutions often have lower
energies than prolate ones. For nuclei with $N > 50$, oblate ground
states are very rare and found only in nuclei very close to shell
magics. The dominance of prolate deformations for $N >50$ may be
attributed to the change of the nature of the major shells from the
harmonic-oscillator shell to the Mayer-Jensen shell.


\begin{center}
\framebox[4cm]{Figure \ref{f_oblpro}}
\end{center}

\section{nucleon skins} \label{s_skin}

As discussed in section~\ref{s_intro}, the greatest advantage of the
mesh representation for this paper is that it is fit to describe skins
and halos.

We choose to define the skin by two conditions:
A point $\bm{r}$ is in the proton or neutron skin when
\begin{equation} \label{eq_skin}
  \rho_{\tau}(\bm{r}) > \frac{3}{4} \rho_{\rm tot} ( \bm{r} )
  \;\;\;\; \mbox{and} \;\;\;\;
  \rho_{\rm tot}(\bm{r}) > \frac{\mbox{0.16 fm}^{-3}}{100},
\end{equation}
where $\tau$ specifies proton or neutron and ``tot'' stands for
proton+neutron. A similar definition was given in Ref.~\cite{FOT93}.
The value of 0.16 fm$^{-3}$ is chosen as a standard density of the
nuclear matter.  Since the factors $\frac{3}{4}$ and $\frac{1}{100}$
in the above definitions could have been differently chosen in rather
arbitrary fashions, the absolute sizes of the skin thickness do not
have a very precise meaning.  Nevertheless, these values are useful to
judge how the nucleon skins develop as the nucleus approaches to the
drip line in the nuclear chart.

Fig.~\ref{f_skin} presents the nucleon skin thicknesses calculated
according to the above definition\footnote{
In the mesh representation, the density is given only at the mesh
points. In order to calculate the skin thickness in the $x$-axis more
accurately than the mesh spacing, first, we interpolate the density at
the points of intersections between the $x$-axis and the mesh planes
$x=(i+\frac{1}{2})a$ using polynomials of order two in $y^2$ and
$z^2$.  Second, interpolations to arbitrary points in the $x$-axis are
done for the logarithm of the density using polynomials of order three
in $x$.  The resulting density in logarithmic plots versus $x$ shows
no artificial ripples due to the interpolations.
}.
We put an open (solid) circle if the proton (neutron) skin exists for
each nucleus.  The diameter of the circle is proportional to the skin
thickness.  Among even-even nuclei inside the proton drip line, only
nine have the proton skin (the heaviest one is $^{46}_{26}$Fe$_{20}$).
Proton skins thicker than 1 fm are found in
$^{12}_{\phantom{0}8}$O$_{4}$ (1.3 fm), $^{6}_{4}$Be$_{2}$ (1.2 fm),
and $^{22}_{14}$Si$_{8}$ (1.1 fm),
all of which are single-magic spherical nuclei.  On the
other hand, the neutron skin has non-zero thickness in 39\% of the
1029 nuclei which we have calculated.  In middle-weight nuclei
$^{134}_{\phantom{0}50}$Sn$_{84}$ and
$^{136}_{\phantom{0}50}$Sn$_{86}$, between which the
present experimental neutron-rich frontier passes\cite{Se94}, the
neutron skin is as thick as 1.02 fm and 1.19 fm, respectively.  The
growth of the skin along isotope and isotone chains is monotonous and
have no irregularity irrespective of proton or neutron and both inside
and outside the proton drip line.


\begin{center}
\framebox[4cm]{Figure \ref{f_skin}}
\end{center}

When the nucleus is deformed, the thickness of the skin depends on the
direction.  (In Fig.~\ref{f_skin}, we take the arithmetic average of
the thicknesses in the $x$-, $y$- and $z$-axes.)  The ratio of the
difference between the maximum and the minimum among the three axes to
the average over the three axes is 8\% for the proton skins and
20\% for the neutron skins on the average.  The relation between the
anisotropy of the skin thickness and the quadrupole deformation
parameter $\delta$ is shown in Fig.~\ref{f_skindef}.  For each ground
or first-excited solution having nucleon skins, a symbol is put at a
point whose abscissa is $\delta$ and its ordinate the skin thickness
in the symmetry axis subtracted by that in the equatorial plane.  The
solid line is the result of a least-square fitting, while two dash
lines correspond to twice as large mean-square deviation as the
minimum value and can be used to judge the quality of the fitting.
For neutron skins, there is a tendency that the skin is thicker in the
symmetry axis than in the equatorial plane for oblate deformations and
vice versa for prolate deformations: The neutron skin tends to make
density more spherical.  For protons, it is an interesting question
whether the Coulomb repulsion between protons can reverse the trend in
such a way that the proton skin promotes deformation.  However, one
cannot see any clear tendency in the right-hand side of
Fig.~\ref{f_skindef}, mainly because the number of points are fewer
(heavy nuclei do not have proton skins).


\begin{center}
\framebox[4cm]{Figure \ref{f_skindef}}
\end{center}

The nucleon halo, being composed of only two nucleons, has much lower
density than the nucleon skin.  We define the halo radius as the
largest $r$ satisfying the following condition,
\begin{equation} \label{eq_halo}
  \rho_{\rm tot}(r) \ge \frac{\mbox{0.16 fm}^{-3}}{10000},
\end{equation}
where $\rho_{\rm tot}(r)$ is the angle-averaged mass density.  Let us
also define the halo thickness as the halo radius subtracted by 1.2
$A^{1/3}$ fm.  According to these definitions, the halo radius
(thickness) in the salient case of $^{11}_{\phantom{0}3}$Li$_{8}$ is
11.2 fm (8.5 fm) while the neutron skin thickness defined by
Eqs.~(\ref{eq_skin}) is only 1.9 fm (from Fig.~4 of Ref.~\cite{TS90}
and Fig.~4 of Ref.~\cite{Sa92}).
In Fig.~\ref{f_sknhal}, we show the relation between the
skin and the halo thicknesses. One can see that the halo grows much
slower than the skin for thin-skin nuclei (i.e.\ near the
$\beta$-stability line) but by far faster for thick-skin nuclei
(i.e.\ near the drip lines).  The maximum value of the halo thickness
is 6.9 fm of $^{50}_{16}$S$_{34}$ (indicated by letter {\bf C} in the
figure) among the nuclei inside the drip lines.  This value is
somewhat smaller than 8.5 fm of $^{11}$Li.  One should note, however,
that such a large halo as in $^{11}$Li can be formed only in a very
restricted interval of the last-occupied nucleon level\cite{Sa92}. It
is possible that more gigantic halos are produced for different force
parameters.


\begin{center}
\framebox[4cm]{Figure \ref{f_sknhal}}
\end{center}

\section{Summary} \label{s_summary}

In this paper we have presented the results of our extensive
Hartree-Fock+BCS calculations with the Skyrme SIII force for 1029
even-even nuclei with $2 \le Z \le 114$ from outside the proton drip
line by several neutrons to beyond the experimental neutron-rich
frontier by a few neutrons.

The single-particle wavefunctions are expressed in a three-dimensional
Cartesian-mesh representation, whose advantages for atomic nuclei have
been discussed.  We have explained the newly developed points of the
method of calculation, including the determination of the pairing
force strengths, the acceleration of the convergence to the HF+BCS
solution using an external quadrupole potential, and the correction of
the total energy for the finite mesh size.

We have compared the calculated ground-state masses with experimental
data and with predictions by other mass models in
section~\ref{s_mass}.  The error from experiment is negative
for most of spherical nuclei, while for deformed nuclei it is
$\sim$ 3 MeV rather independently of the size of deformation.  The
r.m.s.~value of the error for 480 even-even nuclei is 2.2 MeV, which
is much larger than the precision ($\sim$0.5 MeV) of recent
extensively-parameter-fitted mass models like the TUYY and the FRDM.
However, the proton drip line in the nuclear chart is almost identical
to that of the FRDM.  Even the location of the neutron drip line is
not very distant from that of the FRDM and the TUYY.

Deformations are discussed in section~\ref{s_deform}.  Because of the
D$_{\rm 2h}$ symmetry imposed on the solutions, only even electric
multipole moments do not vanish.  We have compared the electric
intrinsic axial quadrupole moments of our solutions with those deduced
from experimental B(E2)$\uparrow$.  The agreements between them are
good except those nuclei with small quadruple moments. After defining
the deformation parameters $a_{lm}$ for the HF+BCS solutions
in terms of multipole moments, we have
plotted the resulting $a_{20}$ and $a_{40}$ in the $(N,Z)$ plane. The
magnitudes of
non-axial deformation parameters $a_{22}$, $a_{42}$, and $a_{44}$ have
turned out very small
($\vert a_{22} \vert$, $\vert a_{42} \vert$ $< 10^{-4}$,
 $\vert a_{44} \vert$ $\sim 10^{-3}$).
The difference of shapes between protons and neutrons have been found
small, too.  Detailed discussions have been presented for
$^{12}$C and $^{80}$Zr.  We have also shown the excitation energy
between the oblate and the prolate solutions and pointed out a clear
difference between below and above the $N=50$ shell magic.

Nucleon skins are discussed in section~\ref{s_skin}. We have shown
that the skin grows monotonously and regularly as nucleons are added
to the nucleus.  On the other hand, the halo grows very slowly except
near the drip lines, where it changes the behavior completely and
expands very rapidly.  It has also been found that the neutron skin
tends to make the density distribution more spherical.

Among the future problems for the extensive mean-field calculations of
the nuclear ground-states properties are the improvements of the
method to determine the pairing force strengths and the extension of
the calculation to the neutron drip line by incorporating the coupling
to the continuum in the pairing channel.  Our final goal is the
comparison of various Skyrme forces in their ability to reproduce the
nuclear ground-state properties globally in the nuclear chart, and
possibly the improvements of the force parameters.

\vspace{\baselineskip}


All the results of the extensive HF+BCS calculations reported in this
paper are available electronically as an
anonymous ftp service on the internet:
$\mbox{\bf nt1.c.u-tokyo.ac.jp}$.  See a file {\bf read.me} in the
home directory for instructions.  The available quantities are the
binding energies, the Fermi levels, the pairing gaps, proton and
neutron moments ($r^2$, $r^2 Y_{2,0}$, $r^2 Y_{2,2}$ $r^4 Y_{4,0}$,
$r^4 Y_{4,2}$, $r^4 Y_{4,4}$), the skin thicknesses, the halo radii,
and the proton/neutron/mass deformation parameters
($a_{20}$, $a_{22}$, $a_{40}$, $a_{42}$, $a_{44}$, $R_0$, $\rho_0$)
of 1029 ground states and 758 first-excited local minima.
Single-particle spectra and density distributions of protons and
neutrons are also obtainable for each solution.  Additional postscript
figures and some FORTRAN source codes to analyze the results are also
provided.


\vspace{\baselineskip}

\noindent {\Large \bf  Acknowledgements}

\vspace{\baselineskip}

The authors thank Dr.~P.~Bonche, Dr.~H.~Flocard, and Dr.~P.-H.~Heenen
for providing the HF+BCS code {\em EV8}.  They are also grateful to
Dr.~T.~Tachibana and Dr.~K.~Oyamatsu for the TUYY mass formula code,
to Dr.~S.~Raman for supplying the computer file of the B(E2)$\uparrow$
table, and to Dr.~P.~M{\"o}ller, Dr.~Dobaczewski, and Dr.~Nazarewicz
for discussions.  This work was financially supported by RCNP, Osaka
University, as RCNP Computational Nuclear Physics Project
(Nos.~94-B-01 and 95-B-01).  Part of the calculations were performed
with a computer VPP500 at RIKEN (Research Institute for Physical and
Chemical Research, Japan).



\newpage


\noindent {\bf TABLES}

\newcounter{tabno}
\begin{list}
{Table \arabic{tabno}. }{\usecounter{tabno}
   \setlength{\labelwidth}{2cm}
   \setlength{\labelsep}{0.5mm}
   \setlength{\leftmargin}{15mm}
   \setlength{\rightmargin}{0mm}
   \setlength{\listparindent}{0mm}
   \setlength{\parsep}{0mm}
   \setlength{\itemsep}{0.5cm}
   \setlength{\topsep}{0.5cm}
}
\baselineskip=\baselineskipTaj
\item \label{t_becorr}
Estimated errors of the total energy due to the finite mesh size of
$a$=1 fm for oblate, spherical, and prolate solutions of five nuclei
in MeV.  The sign of the errors are inverted.  The last column shows
the values given by the correction formula~(\ref{eq_fit5}).  See text
for explanations.
\begin{center}
\begin{tabular}{crrrr}
\hline
   & oblate & spherical & prolate & correction \\
\hline
$^{\phantom{1}72}_{\phantom{1}34}$Se$_{38}$ & 2.9 & 3.1 & 2.8 & 2.83\\
$^{100}_{\phantom{0}40}$Zr$_{60}$           & 4.4 & 4.7 & 4.2 & 4.30\\
$^{130}_{\phantom{0}60}$Nd$_{70}$           & 4.9 & 5.1 & 4.7 & 4.99\\
$^{170}_{\phantom{0}68}$Er$_{102}$          & 7.1 & 7.3 & 6.7 & 7.14\\
$^{240}_{\phantom{0}94}$Pu$_{146}$          &10.1 &10.0 & 9.9 &10.05\\
\hline
\end{tabular}
\end{center}
\vspace*{\baselineskip}
\item \label{t_massdev}
The r.m.s.\ differences of nuclear masses between experiment
and theoretical models and between theoretical models in MeV.  In
parentheses are the number of even-even nuclei to take the mean
value. The FRDM (the ETFSI) does not give masses for $N <8$ or $Z<8$
(for $A<36$), while AW'93 and EV8C do not extend to the neutron drip
line.  See text for explanations.
\begin{center}
\begin{tabular}{lrrrrrrrrrr}
\hline
 & \multicolumn{2}{c}{AW'93} & \multicolumn{2}{c}{TUYY}
 & \multicolumn{2}{c}{FRDM}  & \multicolumn{2}{c}{ETFSI}
 & \multicolumn{2}{c}{EV8C} \\
\hline
TUYY  &0.52&(480)&     &      &     &      &    &      &    &      \\
FRDM  &0.68&(462)& 4.31&(1521)&     &      &    &      &    &      \\
ETFSI &0.74&(430)& 4.27&(1472)& 2.74&(1742)&    &      &    &      \\
EV8C  &2.22&(480)& 2.59& (977)& 2.50& (958)&2.26& (940)&    &      \\
macro &3.55&(480)&17.25&(2228)&16.07&(2246)&8.29&(1895)&5.71&(1029)\\
\hline
\end{tabular}
\end{center}
\vspace*{\baselineskip}
\item \label{t_improv}
Same as in Table~\ref{t_massdev}, but the r.m.s.\ errors have been
decreased by adding Bethe-Weizs{\"a}cker type functions to the
models. See text for explanations.
\begin{center}
\begin{tabular}{lrrrr}
\hline
           &   AW'93  &   TUYY  &   FRDM  & ETFSI  \\
\hline				
TUYY       &   0.50   &         &         &        \\
FRDM       &   0.68   &   1.73  &         &        \\
ETFSI      &   0.68   &   1.53  &   2.10  &        \\
EV8C       &   1.62   &   1.80  &   1.57  &   1.54 \\
\hline
\end{tabular}
\end{center}
\end{list}



\newpage

\noindent {\bf FIGURE CAPTIONS}

\newcounter{figno}
\begin{list}
{Fig. \arabic{figno}. }{\usecounter{figno}
   \setlength{\labelwidth}{1.3cm}
   \setlength{\labelsep}{0.5mm}
   \setlength{\leftmargin}{8.5mm}
   \setlength{\rightmargin}{0mm}
   \setlength{\listparindent}{0mm}
   \setlength{\parsep}{0mm}
   \setlength{\itemsep}{0.5cm}
   \setlength{\topsep}{0.5cm}
}
\baselineskip=\baselineskipTaj
\item \label{f_conv}
Imaginary-time evolution of the deformation parameter $\delta$ (top
portion),
the maximum energy spreading of the single-particle states
$\Delta \epsilon$ (bottom-left portion), and the total energy relative
to the convergent value $\Delta E$ (bottom-right portion) for
$^{156}$Er. The initial single-particle wavefunctions are taken from
those of the HF+BCS solution for $^{158}$Er.  The abscissae of the
three graphs represent the number of steps of the imaginary-time
evolution.  The dot curves represent the history of a free evolution
while the solid curves are obtained by switching on an external
quadrupole potential between time steps 213 and 291 to accelerate the
convergence.  The quantity $\delta$ is calculated in every step, while
$\Delta \epsilon$ and $\Delta E$ are in every 25 steps.
\item \label{f_masx}
Nuclear masses calculated with the HF+BCS with SIII corrected
according to Eq.~(\ref{eq_fit5}).  The smooth part of
Bethe-Weizs{\"a}cker form has been subtracted.  The solid
staircase-like lines designate two-nucleon drip lines from our
calculations. The dash lines are those from the FRDM\cite{MNM94}.
\item \label{f_masxdf}
Error of the nuclear masses calculated with the HF+BCS with SIII
corrected according to Eq.~(\ref{eq_fit5}). The grid indicates the
locations of the magic numbers.
\item \label{f_qq}
Comparison between the experimental and the calculated intrinsic
quadrupole moments in unit of barn.  For each nucleus whose
B(E2)$\uparrow$ is given in Ref.~\cite{RMM87} (289 even-even nuclei
ranging over 4 $\le Z \le$ 98), a dot is put at a point whose abscissa
is equal to the experimental value and its ordinate to the value
calculated with the HF+BCS with SIII. The diagonal line is drawn so
that one can see easily the quality of the agreement. See text for
explanations.
\item \label{f_a20}
The quadrupole deformation parameter $a_{20}$ of the ground state
solutions of the HF+BCS with SIII for even-even nuclei. Prolate
(oblate) nuclei are designated with open (solid) circles whose
diameter is proportional to $\vert a_{20} \vert$. The staircase-like
lines represent two-nucleon drip lines from our calculations.
\item \label{f_a40}
Same as in Fig.~\ref{f_a20}, but for the hexadecapole deformation
parameter $a_{40}$.  The grid indicates the locations of the magic
numbers.
\item \label{f_zrpes}
The potential energy curves for $^{80}$Zr obtained by solving the
HF+BCS equation with constraint on the mass quadrupole moment $Q_z$.
The abscissa is the deformation parameter $\delta$, while the ordinate
is the energy without correction for the finite-mesh-size effect.  The
solid and the dot curves are calculated with the SIII force, the
former with the standard-strength and the latter with a weaker paring
correlation.  The dash curve is calculated with the SkM$^{\ast}$
force, while the dot-dash curve with the SGII force (vertically
shifted by 15 MeV).
\item \label{f_oblpro}
The energy difference between the oblate and the prolate solutions of
the HF+BCS with SIII.  The abscissa is the neutron number $N$, while
the ordinate is the energy of the oblate solution subtracted by that
of the prolate solution.  A circle is put only when both solutions
exist for each nucleus.  The circles belonging to the same isotope
chain are connected by a line to guide the eyes. The atomic number $Z$
can be known from the direction of the hand in each circle.
\item \label{f_skin}
The thickness of the proton and the neutron skins of the ground states
of the HF+BCS with SIII.  The proton (neutron) skin thickness is
proportional to the diameter of open (solid) circles.  The
staircase-like lines represent two-nucleon drip lines from our
calculations.
\item \label{f_skindef}
The dependence of the anisotropy of the proton and the neutron skins
on the quadrupole deformation parameter $\delta$.  The circle (plus)
symbol is used for nuclei with $A \le 100$ ($>100$).  See text for
explanations.
\item \label{f_sknhal}
The relation between the thicknesses of halos and skins.  For each of
the ground and first-excited solutions of even-even nuclei, a plus
symbol is plotted when the skin is of protons or there is no skins,
while an x symbol is put when the skin is of neutrons.  The plus
symbols indicated by letters {\bf A} and {\bf B} correspond to
$^{22}_{16}$S$_{6}$ and $^{20}_{14}$Si$_{6}$, respectively, which are
outside the proton drip line. The x symbol indicated by letter {\bf C}
corresponds to $^{50}_{16}$S$_{34}$.
\end{list}

\end{document}